\documentclass[letter]{aa} 

\usepackage{graphicx}
\usepackage{natbib}
\usepackage{txfonts}
\newcommand\xmm{{\sl XMM-Newton}}
\newcommand\chandra{{\sl Chandra}}
\newcommand\nustar{{\sl NuSTAR}}

\newcommand\kms{\rm{km\,s^{-1}}}
\newcommand{\fluxcgs}{erg~s$^{-1}$~cm$^{-2}$}
\newcommand{\lumcgs}{erg~s$^{-1}$}
\newcommand{\lum}{$L_{\scriptstyle \rm X}$}
\newcommand{\luv}{$L_{\scriptstyle \rm UV}$}
\newcommand{\lmir}{$L_{\scriptstyle \rm MIR}$}
\newcommand{\lognh}{{\rm log} (N_{\rm H}/{\rm cm^{-2}})}

\newcommand{\nh}{cm$^{-2}$}
\newcommand{\nhi}{\rm cm^{-2}}
\newcommand{\nhsym}{N_{\rm H}}

\newcommand{\lumh}{L_{\rm 2-10}}
\newcommand{\luvmon}{L_{\rm 2500\rm\AA}}
\newcommand{\lmirmon}{L_{\rm 6\mu m}}

\newcommand{\lbol}{L_{\scriptstyle \rm bol}}
\newcommand{\ledd}{\lambda_{\rm Edd}}

\newcommand{\aox}{\alpha_{\scriptscriptstyle \rm OX}}

\def\vciv{$v_{\scriptscriptstyle \rm CIV}$}

\begin{document}

   \title{The WISSH quasars project VII. The impact of extreme radiative field in the accretion disc and X-ray corona interplay}

   \titlerunning{Impact of extreme radiative field in quasar  disc and corona}
   
   \author{L. Zappacosta
          \inst{1}
          \and E. Piconcelli\inst{1}
          \and M. Giustini\inst{2}
          \and G. Vietri\inst{3}
          \and F. Duras\inst{4}
          \and G. Miniutti\inst{2}
          \and M. Bischetti\inst{1}
          \and A. Bongiorno\inst{1}
          \and M. Brusa\inst{5,6}
          \and M. Chiaberge\inst{7}
          \and A. Comastri\inst{6}
          \and C. Feruglio\inst{8}
          \and A. Luminari\inst{9,1}
          \and A. Marconi\inst{10,11}
          \and C. Ricci\inst{12,13}
          \and C. Vignali\inst{5,6}
          \and F. Fiore\inst{8}
          }

   \institute{INAF - Osservatorio Astronomico di Roma, via di Frascati 33, 00078 Monte Porzio Catone, Italy
     \and  Centro de Astrobiolog\'ia (CSIC-INTA), Dep. de Astrof\'isica; Camino 
Bajo del Castillo s/n, Villanueva de la Ca\~nada, E-28692 Madrid, 
Spain
     \and INAF - IASF Milano, via A. Corti 12, 20133 Milano, Italy
     \and Aix Marseille Univ, CNRS, CNES, LAM, Marseille, France
     \and Dipartimento di Fisica e Astronomia, Università degli Studi di Bologna, via Gobetti 93/2, 40129 Bologna, Italy
     \and INAF – Osservatorio di Astrofisica e Scienza dello Spazio di Bologna, Via Gobetti 93/3, 40129 Bologna, Italy
     \and Department of Physics and Astronomy, Bloomberg Center, Johns Hopkins University, Baltimore, MD 21218, USA
     \and INAF – Osservatorio Astronomico di Trieste, Via G. Tiepolo 11, I-34143, Trieste, Italy
     \and Department of Physics, University of Rome “Tor Vergata”, Via della Ricerca Scientifica 1, I-00133 Rome, Italy
     \and Dipartimento di Fisica e Astronomia, Universit\`a di Firenze, Via G. Sansone 1, I-50019, Sesto Fiorentino (Firenze), Italy
     \and INAF-Osservatorio Astrofisico di Arcetri, Largo E. Fermi 5, 50125, Firenze, Italy
     \and N\'ucleo de Astronom\'ia de la Facultad de Ingenier\'ia, Universidad Diego Portales, Av. Ej\'ercito Libertador 441, Santiago, Chile
     \and     Kavli Institute for Astronomy and Astrophysics, Peking University, Beijing 100871, China
}

   \abstract
       {Hyper-luminous quasars ($\lbol\gtrsim10^{47}$~\lumcgs) are ideal laboratories to study the interaction and impact of the extreme radiative field and the most powerful winds in the active galactic nuclei (AGN) nuclear regions. They typically exhibit low coronal X-ray luminosity (\lum) compared to the ultraviolet (UV) and mid-infrared (MIR) radiative outputs (\luv\ and \lmir); a non-negligible fraction of them report even $\sim$1~dex weaker \lum\ compared to the prediction of the well established \lum-\luv\ and \lum-\lmir\ relations followed by the bulk of the AGN population. In our WISE/SDSS-selected Hyper-luminous (WISSH) $z=2-4$ broad-line  quasar sample, we report on the discovery of a dependence between the intrinsic 2-10~keV luminosity ($\lumh$) and the blueshifted velocity of the CIV emission line (\vciv) that is indicative of accretion disc winds. In particular, sources with the fastest winds (\vciv$\gtrsim3000~\kms$) possess $\sim$0.5-1~dex lower $\lumh$ than sources with negligible \vciv.  No similar dependence is found on \luv, \lmir, $\lbol$, the photon index, or the absorption column density.
We interpret these findings in the context of accretion disc wind models. Both magnetohydrodynamic and line-driven models can qualitatively explain the reported relations as a consequence of X-ray shielding from the inner wind regions. In case of line-driven winds, the launch of fast winds is favoured by a reduced X-ray emission, and we speculate that these winds may play a role in directly limiting the coronal hard X-ray production.
       }


   \maketitle
%

\section{Introduction}
The most luminous active galactic nuclei (AGN) are expected to exhibit the strongest and clear-cut manifestations of winds \citep[][]{M2008,FGQ2012}.
Indeed, the fastest and most energetic winds have been reported in hyper-luminous quasars \citep[i.e. with a bolometric luminosity of $\lbol\gtrsim10^{47}~\rm erg s^{-1}$;][]{W2011,F2017,V2018,M2019,P2019}. Luminous quasars are typically characterised by their low coronal X-ray luminosity (\lum) compared to the disc's ultraviolet (UV) and larger-scale dust-reprocessed mid-infrared (MIR) luminosities, which are represented as \luv\ and \lmir, respectively. This characterisation is usually performed by adopting the $\aox$\footnote{$\aox$ is the X-ray to optical spectral index between the rest-frame luminosities at $2500~\rm\AA$ and 2~keV; i.e. $\aox=0.3838\, \log(L_{\rm 2 keV}/L_{\rm 2500~\rm\AA})$}  and $L_{\rm MIR}/L_{\rm X}$ parameters \citep[e.g.][]{V2003,J2007,LR2016,St2015,M2017,Ch2017}. 
Past studies on large quasar samples over a wide luminosity range ($\lbol\approx10^{45}-10^{48}$~\lumcgs) have also reported indications of a further weak dependence of the $\aox$ parameter on the velocities of the broad emission line (BEL) winds, which are parametrised by the CIV blueshift, that is, negative velocity shifts (\vciv) of the CIV emission line \citep[e.g.][]{R2011,K2011,V2018}. However, once the $\aox$ luminosity dependence is removed by adopting $\Delta\aox$, that is, the difference between the measured $\aox$ and the one predicted by the $\aox-L_{\rm 2500~\rm\AA}$ relation \citep[e.g.][]{J2007}, the dependence with \vciv\ is much less significant \citep[e.g.][]{G2008,N2018,T2019}.
It is important to notice that these dependences on \vciv\ may be regarded as representative for the bulk of the quasar population since CIV BEL winds are a common feature in quasars \citep[e.g.][]{Sh2011}. A similar dependence has also been reported for the maximum velocity measured in line-of-sight detected winds, such as those reported in the broad-absorption line (BAL) quasars \citep[e.g.][]{G2006,G2009,S2011}. 
Nonetheless, a peculiar category of $z=2-3$ optically luminous quasars selected to have weak broad emission lines (with a rest-frame equivalent width of the CIV, $REW_{\rm CIV}<15~\rm\AA$) and fast BEL winds \citep[\vciv$\lesssim~-2000~\kms$][]{R2011}, has mostly revealed 1-2~dex weaker  X-ray emission \citep{G2008,W2011} compared to the $\aox-L_{\rm 2500~\rm\AA}$ expectations.

\begin{figure*}[t!]
   \begin{center}
     \includegraphics[height=0.32\textwidth,angle=0]{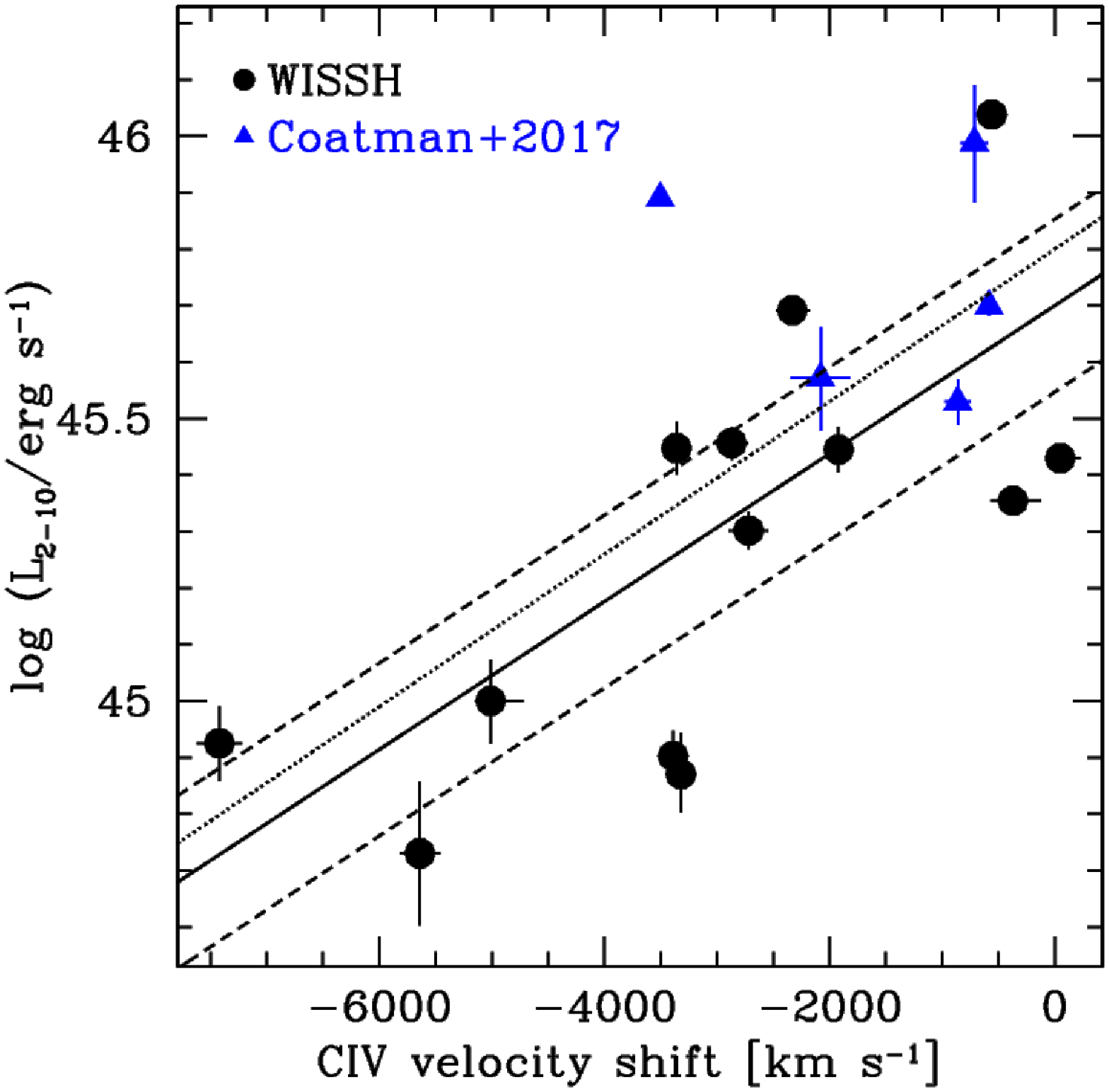}
     \includegraphics[height=0.32\textwidth,angle=0]{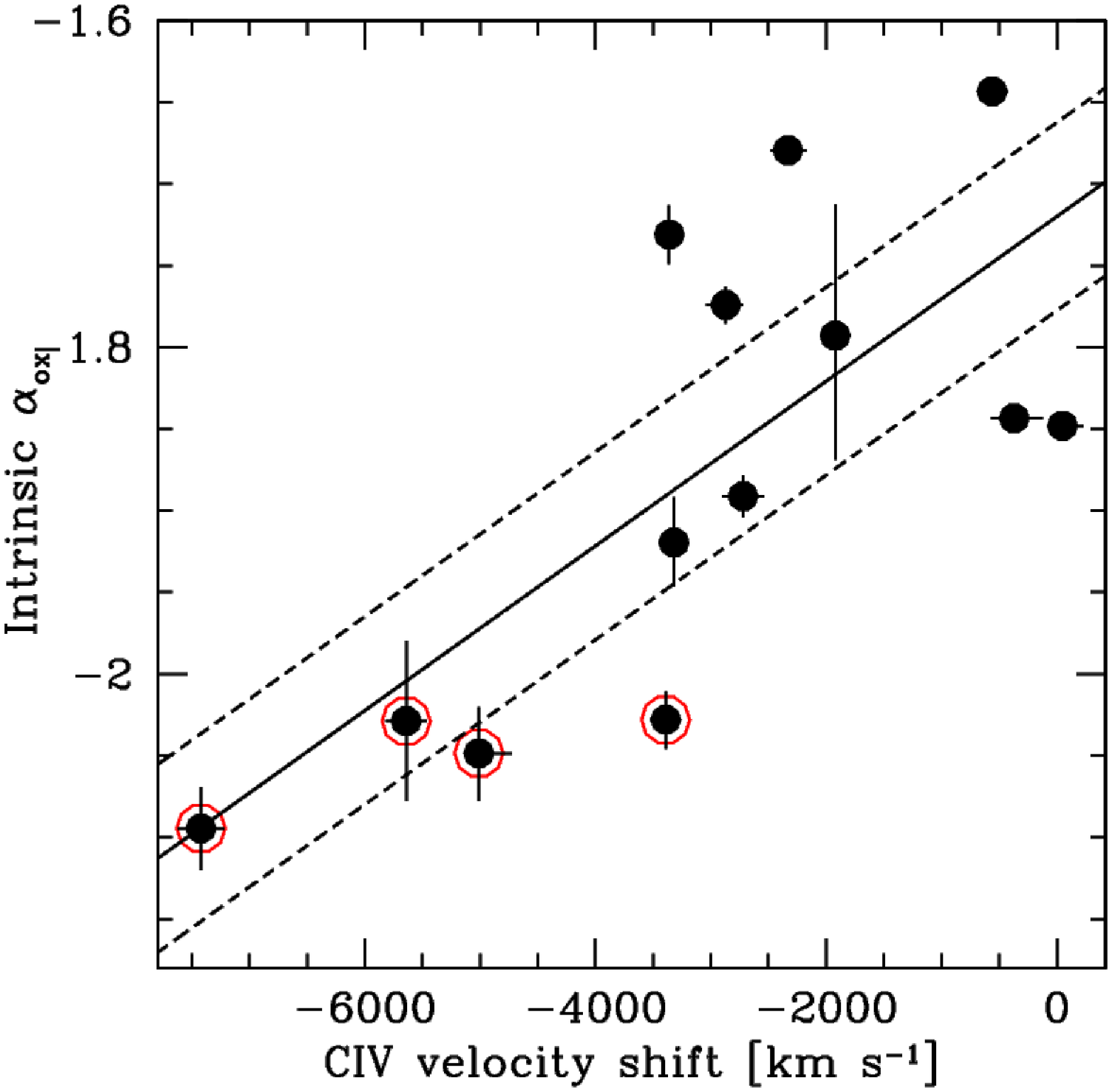}
     \includegraphics[height=0.32\textwidth,angle=0]{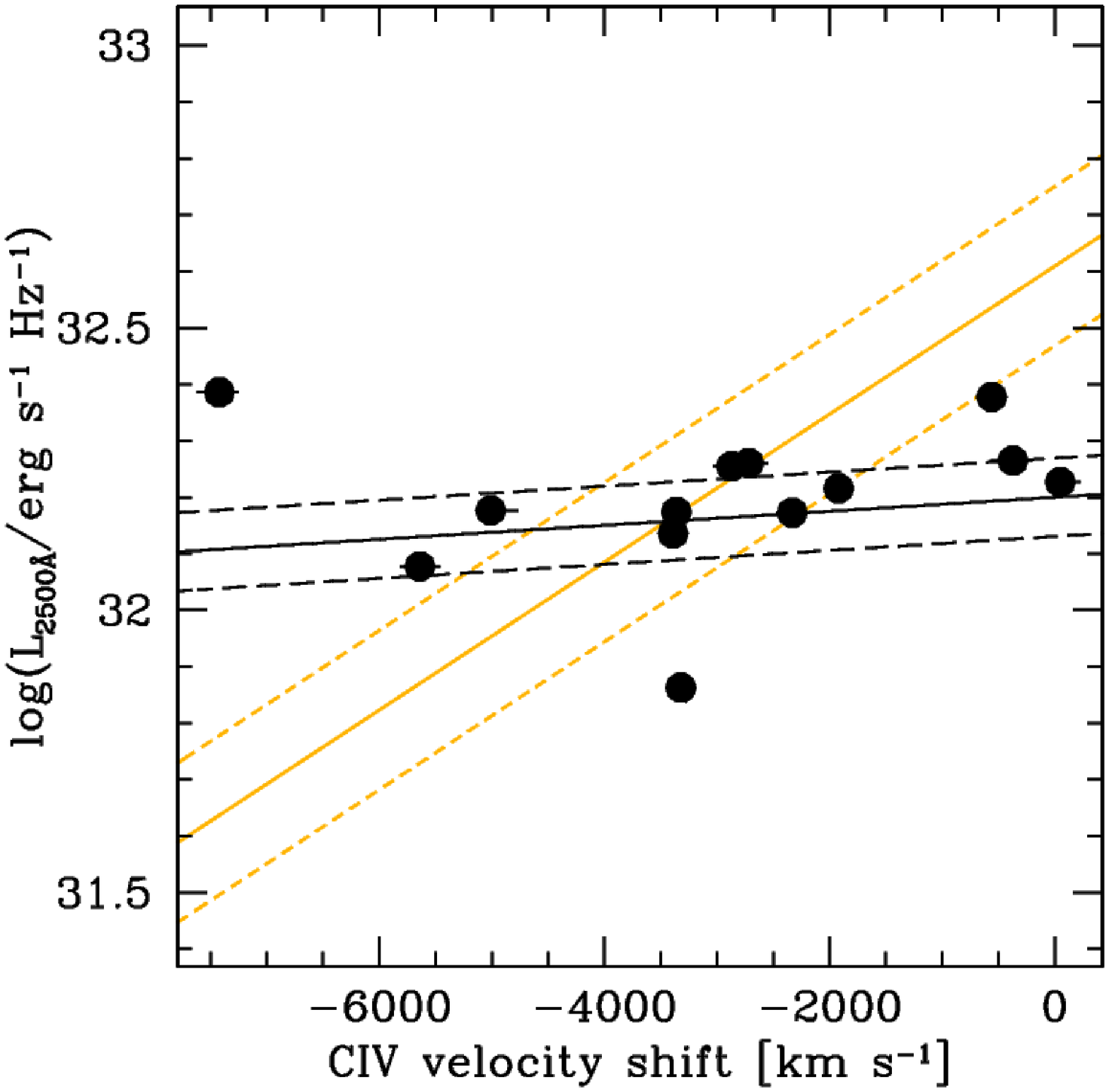}
   \end{center}
   \caption{Left panel: $\lumh$ as a function of \vciv\ for the WISSH quasars (black circles). We also report hyper-luminous quasars at $z=2-3$  (C17 sample, blue triangles).  Middle panel: Intrinsic $\aox$ as a function of \vciv\ for the WISSH quasars. Red circles indicate X-ray weak sources (see Sect.~\ref{results} for details). Right panel: $L_{\rm 2500~\rm\AA}$ as a function of \vciv. Solid and dashed lines report the best-fit linear relation and $1\sigma$ uncertainties on the normalisation for the WISSH-only relations. Dotted line in the left panel reports the best-fit relation with the addition of the C17 sample.  Orange lines in the right panel report the best-fit $\lumh$-\vciv\ relation normalised to the mean values of $L_{\rm 2500~\rm\AA}$.} 
   \label{lxciv}
\end{figure*}

Sources with \vciv$\lesssim~-2000~\kms$ and $REW_{\rm CIV}\lesssim20~\rm\AA$ are reported in our sample of 86 broad-line unlensed highly accreting ($\ledd>0.4$) MIR and optically bright $z=2-4$ hyper-luminous quasars. These sources were selected to be the MIR-brightest WISE/SDSS quasars with $z>1.5$ and a flux density of $S_{22\mu m}> 3~\rm mJy$ \citep[WISSH;][]{BP2017}. The WISSH quasars exhibit widespread evidence of winds at all scales from nuclear BAL \citep{B2019} to ionised $\rm [OIII]/Ly\,\alpha$ galactic- and circumgalactic-scale outflows \citep[][]{BP2017,Tr2020}. In particular, in \citet{V2018} we report a surprisingly high fraction ($\sim70\%$) of sources with  weak  UV and optical BEL spectra (e.g. $REW_{\rm CIV}\lesssim20~\rm\AA$) as well as extreme CIV blueshifts (\vciv$<-2000~\kms$) in a sub-sample of the WISSH quasars. Furthermore, \citet{M2017} found a large spread in the \lum\  with a non-negligible fraction that have $\sim$0.5-1~dex fainter values than the average.

In this Letter, we explore the relation between the extreme radiative field of the hyper-luminous WISSH quasars and their X-ray coronal properties, and we interpret it in the context of accretion disc wind scenarios. 
We adopt a $\Lambda$CDM cosmology with $\Omega_\Lambda=0.73$ and $H_0=70\,\rm  km\, s^{-1} Mpc^{-1}$ throughout the paper. Errors as well as upper and lower limits are quoted at $68.3\%$ and $90\%$ confidence levels, respectively.

\section{Sample presentation and X-ray data reduction and analysis}
In this work we consider the radio-quiet  hyper-luminous WISSH sources  with (i) reported \vciv\ measures that are relative to their systemic redshift and (ii) available X-ray data. The selected sample of thirteen sources is reported in Table~\ref{wisshprop} and was mainly drawn from the \citet{V2018} WISSH sub-sample of 18 quasars for which CIV emission line properties were derived. We also complement our work by including five Type~1 radio-quiet hyper-luminous sources at similar $z$ with published \vciv\ \citep[][ hereafter C17; see Table~\ref{wisshprop}]{C2017} and available X-ray archived data. Further details on the sample selection are reported in Appendix~\ref{sampleselection}.

We consider both \chandra\ and \xmm\ observations that are available for each source (see Table~\ref{xraysample}).
For each dataset, we performed standard data reduction as detailed in Appendix~\ref{xrayreduction}. 
The X-ray spectral modelling was performed in the 0.2-10~keV and 0.3-8~keV bands for \xmm\ and \chandra, respectively. We employed an intrinsically absorbed power-law model, which was further modified by the Galactic absorption.   Further details on the modelling and the derived parameters are reported in Appendix~\ref{xrayanalysis} and Table~\ref{wisshxray}.

\section{Results}\label{results}
Fig.~\ref{lxciv} shows the unabsorbed (i.e. intrinsic) 2-10~keV luminosity ($\lumh$; left panel) and the X-ray unabsorbed and UV de-extincted $\aox$ (middle panel) as a function of \vciv\ for the WISSH quasars (black). Both quantities strongly correlate  with \vciv\footnote{See Table~\ref{wisshprop} for its definition.} with the Spearman correlation coefficient $\rho\approx0.6$ and the two-sided null-hypothesis probability of $p\approx0.02$ (see Table~\ref{corr_anal}). This is a consequence of the lack of significant correlation between $\luvmon$ and \vciv\ (right panel; $\rho=0.32$ and $p\approx0.29$) and the limited dispersion of $\luvmon$. Hence the sources with the largest negative \vciv\ (i.e. larger blueshifts) are $0.5-1$~dex X-ray weaker and exhibit steeper $\aox$ than the sources with the lowest \vciv.  

\begin{figure}[t]
   \begin{center}
     \includegraphics[width=0.49\textwidth,angle=0]{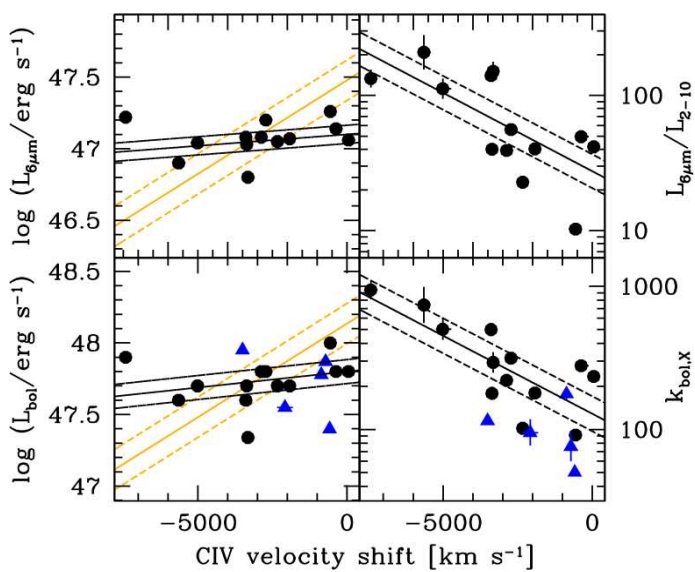}
   \end{center}
   \caption{Top-panels: $\log L_{6\mu m}$ (left) and $L_{\rm 6\mu m}/L_{\rm 2-10}$ (right) as a function of \vciv. Bottom panels: $\log \lbol$ (left) and $k_{\rm bol,X}$ (right) as a function of \vciv. The meaning of the data points and lines is the same as in Fig.~\ref{lxciv}.}
   \label{lumsciv}
\end{figure}

No significant correlation between \vciv\ and $\lbol$ and $\lmirmon$ is found (see left panels in Fig.~\ref{lumsciv} and  Table~\ref{corr_anal}). 
Therefore their ratio with $\lumh$, that is, the X-ray bolometric correction $k_{\rm bol,X}=\lbol/\lumh$ and $L_{\rm 6\mu m}/L_{\rm 2-10}$, are strongly dependent on \vciv\ (see right panels in Fig.~\ref{lumsciv} and  Table~\ref{corr_anal}). The inclusion of the hyper-luminous quasars from C17 in the relations involving $\lumh$, $\lbol,$ and $k_{\rm bol,X}$  further confirms the strength and significance of the dependence with \vciv. We note that \citet{V2018} and C17 adopted slightly different definitions of \vciv\ (see notes on Table~\ref{wisshprop}). This difference does not change our result.
Indeed, for the two WISSH quasars reported in both samples (i.e. J1106+6400 and J1201+1206), the \vciv\ reported by C17 are $400-500~\rm km~s^{-1}$ larger than the values reported by \citet{V2018}. This small systematic offset makes little difference in our correlations and if we correct the \vciv\ of the C17 sub-sample by $500~\rm km~s^{-1}$, we obtain a slightly stronger $\lumh-$\vciv\ correlation with $\rho=0.63$ and $p=0.005$.\footnote{We mention that the relation between $\lumh$ and the CIV BEL wind velocity is still in place ($\rho\approx0.77$ and $p=0.003$) even if we replace \vciv, the peak of the CIV emission line, with the velocity of the CIV outflow component as estimated in the two component CIV spectral modelling by \citet{V2018}.}

The $\Delta\aox$ values (based on the $\aox-\luvmon$ relation from \citealt{J2007}) appear to be strongly dependent on \vciv\ (see Table~\ref{corr_anal}).
Similar to past works, we also calculated $\aox^{pow}$, that is, we derived it by performing spectral fitting with an unabsorbed power-law model for the 2~keV luminosity estimation (see Appendix~\ref{xrayanalysis}) and the relative $\Delta\aox^{pow}$(see Table~\ref{wisshxray}). Both quantities exhibit a strong and significant correlation with \vciv, which is similar to $\aox$ and $\Delta\aox$. We find that four and five sources ($\sim30-40$\% of the considered WISSH sub-sample) are X-ray weak (red circles in Fig.~\ref{lxciv}), as they have $\Delta\aox$ and $\Delta\aox^{pow}$ values that are less than $-0.2$ (a typically adopted threshold value, e.g. \citet[][]{L2015}, used to identify X-ray weak sources), respectively.
We notice that, given the small number of sources and the sample selection function (i.e. only X-ray detected sources for which basic X-ray spectral analysis can be performed), the fraction of X-ray weak sources must be considered with caution, although see \citet{N2019} for similar results.  

For all of the significant relations (i.e. with $p\lesssim0.02$), we also computed $\rho$ and account for the uncertainties in the measurements. We generated 10000 random realisations of the sample by Gaussian distributing each quantity that was considered according to its best-fit value and error. We obtain $\rho_{sim}$ and its uncertainty by adopting the mean value and standard deviation of the distribution of $\rho$ for the simulated datasets. We also calculated $p_{sim,90}$, the 90\% percentile on the distribution of the p values. The derived $\rho_{sim}$ and $p_{sim,90}$ confirm the significance of the relations. 
For all quantities, we derived linear relations with \vciv\ by employing the BCES(Y|X) method \citep[][]{AB1996}. We show them in Fig.~\ref{lxciv} and Fig~\ref{lumsciv}.
Table~\ref{corr_anal}  reports $\rho$, p, $\rho_{sim}$, $p_{sim,90}$, and the slope ($\alpha$) and y-intercept ($\beta$) of the linear relation.

As for the derived X-ray spectral parameters, we do not report a significant dependence of the photon index ($\Gamma$) on \vciv\ (see Table~\ref{corr_anal}), but it is important to bear in mind that we have weak constraints on $\Gamma$ for the X-ray weaker sources (see top panel of Fig.~\ref{gnhshift}). As for the column density ($\nhsym$), we find the WISSH quasars to exhibit moderate values of absorption which are compatible with their Type~1 nature. Indeed seven sources have measured $\nhsym\lesssim10^{23}~\rm cm^{-2}$ with a mean $\lognh\approx22.3$. The other sources have upper limits in the range of $\lognh=21-23.7$, with the largest values mainly driven by poor statistics in the spectra ($\lesssim$50~net-counts). The lower panel of Fig.~\ref{gnhshift} reports $\nhsym$ as a function of \vciv. We computed the Spearman's rank correlation coefficient, which accounts for upper limits\footnote{We used the ASURV package v.~1.2 to account for censored data \citep{asurv,FN1985,IFN1986}.}. We find a weak but not significant anti-correlation  both including or excluding the C17 data (see Table~\ref{corr_anal}).

Finally we mention a lack of correlation between $\ledd$ \citep[as reported in][]{V2018} and \vciv\ for the WISSH quasars. Indeed, all of these sources shine at $\ledd\approx 1$ and the reported large uncertainties on the $\ledd$, which are mainly driven by the constraints on the supermassive black-hole (SMBH) mass \citep[see ][]{V2018}, prevent an accurate investigation of a possible trend.

\section{Discussion}
We report a relation between $\lumh$ and $\aox$ with \vciv\ in a sample of
MIR and optically-selected hyper-luminous quasars, and we also note the lack of a similar dependence in the UV, MIR, and in $\lbol$. 
The use of good-quality X-ray data and the well-defined quasar selection, resulting in a narrow $\lbol$ ($\approx \rm L_{UV}$) range, allows one to measure a marked  $\aox$-\vciv\ ($\Delta\aox$-\vciv) dependence \citep[i.e. stronger in terms of the correlation coefficient than those reported to date in samples probing the bulk of the quasar population;][]{R2011,V2018,T2019} and reveals a clear-cut dependence on $\lumh$. This suggests that at these luminosity regimes, $\lumh$ is the main driver of the $\aox$-\vciv\ relation. We notice that the large average $\langle$\vciv$\rangle$ for our quasars of $\sim-2900~\kms$ is in agreement with the increasing trend of the \vciv$\,\propto \lbol^{0.28}$  relation reported, for example in \citet{V2018}, for an extended sample spanning more than three decades in $\lbol$ (see also \citealt{T2019} for a similar result).  Accordingly, the lack of a significant correlation for $\log \lbol$-\vciv\ in our sample is the result of the restricted luminosity range spanned by our quasars.

\setlength{\tabcolsep}{4pt}
\begin{table*}
  \caption{Correlation and linear regression coefficients}
  \label{corr_anal}
  \begin{center}
    \begin{tabular}{lccccccc}
      \hline
      \hline
      Relation          &  sample$^{a}$  &  $\rho$          &     $p^b$          &  $\rho_{sim}^{c}$    &     $p_{sim,90}^{d}$         &    $\alpha^{e}$          & $\beta$          \\
\hline                                                                                                                    
      $\log \lumh$--\vciv   &W        & 0.63             &     0.0210           & 0.68$\pm$0.06    &     0.0340         &   $0.13\pm0.04$    &  $45.7\pm0.2$       \\
                              &W+C17    & 0.53             &     0.0244           & 0.56$\pm$0.04    &     0.0346         &   $0.14\pm0.04$    &  $45.8\pm0.1$        \\
      $\log L_{2500}$--\vciv   &W        & 0.32             &     0.2872           &      --          &       --           &   $0.01\pm0.03$    &  $32.2\pm0.1$       \\
      $\log L_{6\mu m}$--\vciv &W        & 0.27             &     0.3680           &      --          &       --           &   $0.01\pm0.02$    &  $47.1\pm0.1$       \\
      $\log \lbol$--\vciv    &W        & 0.38             &     0.1976           &      --          &       --           &   $0.02\pm0.02$    &  $47.8\pm0.1$       \\
                              &W+C17    & 0.15             &     0.5546           &      --          &       --           &   $0.00\pm0.02$    &  $47.7\pm0.1$      \\
      $k_{\rm bol,X}$--\vciv       &W        & -0.69            &     0.0111           & -0.69$\pm$0.05   &     0.0269         &   $-0.11\pm0.02$   &  $2.1 \pm 0.1$       \\
                              &W+C17    & -0.61            &     0.0087           & -0.61$\pm$0.04   &     0.0163         &   $-0.13\pm0.03$   &  $2.0 \pm 0.1$       \\
      $L_{6\mu m}/L_{2-10}$--\vciv &W     &  -0.67           &     0.0150           & -0.71$\pm$0.05   &     0.0223         &   $-0.11\pm0.02$  &  $1.5\pm0.1$       \\
      $\alpha_{\rm ox}$--\vciv     &W        &  0.68           &      0.0139           & 0.68$\pm$0.04    &      0.0252       &   $0.05\pm0.01$  &  $-1.7\pm0.1$       \\
      $\Delta\alpha_{\rm ox}$--\vciv  &W     &  0.68           &      0.0139           & 0.68$\pm$0.04    &      0.0253       &   $0.05\pm0.01$    &  $0.08\pm0.05$       \\
      $\alpha_{\rm ox}^{pow}$--\vciv &W       &  0.69           &      0.0094           & 0.71$\pm$0.04    &      0.0171       &   $0.08\pm0.01$  &  $-1.7\pm0.1$       \\
      $\Delta\alpha_{\rm ox}^{pow}$--\vciv &W &  0.71           &      0.0088           & 0.71$\pm$0.03    &      0.0160       &   $0.08\pm0.01$  &  $0.13\pm0.06$       \\
      $\Gamma$--\vciv         &W        &   0.24           &     0.4258           &      --          &     --            &   $0.03\pm0.04$      &  $1.95\pm0.11$       \\
                              &W+C17    &   0.28           &     0.2564           &      --          &     --            &   $0.04\pm0.03$      &  $2.01\pm0.08$       \\
      $\log \nhsym$--\vciv    &  W      &    -0.14$^g$    &   0.6295$^g$         &     --            &      --          &   $-0.10\pm0.13$$^h$      &  $ 21.5\pm0.4$$^h$       \\
                              &  W+C17  &    -0.07$^g$    &   0.7645$^g$         &     --            &      --          &   $-0.12\pm0.14$$^h$      &  $ 21.2\pm0.5$$^h$       \\
\hline
    \end{tabular}
  \end{center}
  \tablefoot{$^{a}$: W (WISSH), W+C17 (WISSH + C17 sample) ; $^{b}$: null-hypothesis probability; $^{c}$: mean and standard deviation of the $\rho$ accounting through simulations for errors in the two quantities (see text for details). This has been computed only for the relations with $p\lesssim0.02$; $^{d}$: 90\% percentile of the distribution of the $p$ values obtained from the simulated random datasets (see text for details). This has been computed only for the relations with $p\lesssim0.02$; $^{e}$: units of $10^{-3}$; $^g$: generalised Spearman’s $\rho$ computed with the survival analysis package ASURV v.~1.2; $^h$: linear regression based on EM algorithm with normal distribution computed with the survival analysis package ASURV v.~1.2. The functional form for a $Y$$-$$X$ relation is $Y=\alpha\times X + \beta$.}  
\end{table*}

Theoretical and observational arguments suggest that BEL winds in AGN 
are produced at accretion disc scales \citep[e.g.][]{E2000}, where the AGN luminosity output is large in the UV and X-ray bands.
Accretion disc winds in AGN can be sustained by either magnetic or 
radiative forces;
the observational results of this Letter imply that regardless of the 
driving mechanism of the disc wind, 
the fastest UV winds appear in the X-ray weakest sources. 

In magnetohydrodynamic (MHD)-driven scenarios, the presence of the disc wind does not depend on the X-ray radiative output; however the existence of observable ions does.
Indeed, a general correlation between $\aox$ and, for example, \vciv\ 
is expected to exist from simple photoionisation arguments.
A larger X-ray luminosity is generally effective 
in stripping the bound electrons off the CIV atoms up to large disc 
radii, allowing for the formation of low-velocity winds. Lower X-ray luminosities instead allow for the presence of CIV at distances closer to the SMBH where larger terminal velocities are expected. In particular,
in order to not get over-ionised, the UV wind must be accompanied
by an inner shield of partially ionised gas absorbing the X-ray flux, which
is itself part of the wind driven by the magnetic forces \citep[see e.g. ][]{F2010}. 

In radiation-driven scenarios, the wind is driven by radiation 
pressure on spectral lines, and the relative contribution of the UV 
and X-ray emission is instead very important in determining the 
existence of the wind itself \citep{M1995,P2000,PK2004}. In particular, for a given UV luminosity, a weak X-ray emission (i.e. steep $\aox$) is crucial for the existence of fast radiation-driven winds.
Indeed, in the inner disc regions, the gas opacity to UV transitions drops as the atoms are over-ionised by an intense X-ray flux. In this case, the wind cannot be efficiently accelerated beyond escape velocity and eventually it falls back to the disc as a failed wind (FW).
The FW contains clumps of dense gas in the proximity of the X-ray 
emitting corona.
The FW  effectively shields the gas that is located
farther away from the ionising X-ray photons, and it allows for the acceleration of disc
material at all radii where the radiation pressure is high enough to
overcome the gravitational pull of the SMBH \citep{P2000,PK2004,RE2010}.

A high X-ray flux would produce a vast inner zone
of FW and would allow for the launch of disc winds only at large radii. Conversely,
a lower X-ray flux would  produce a reduced inner FW zone, and would allow  
for the formation of disc winds on scales that are closer to the SMBH.  As
the terminal velocity of the wind is inversely proportional to its
launching distance from the central SMBH, a lower X-ray emission
favours, in general, the launch of faster radiation-driven accretion
disc winds, compared to a higher X-ray emission \citep[see e.g. ][ for a recent review]{GP2019}.

Interestingly, because of their high density, FW clumps can further cause an
efficient cooling of the corona via bremsstrahlung emission.\ This, therefore,
leads to a weakening (quenching) of the inverse Compton X-ray emission \citep{Pr2005,LD2014}.

\begin{figure}[t]
   \begin{center}
     \includegraphics[height=0.45\textwidth,angle=0]{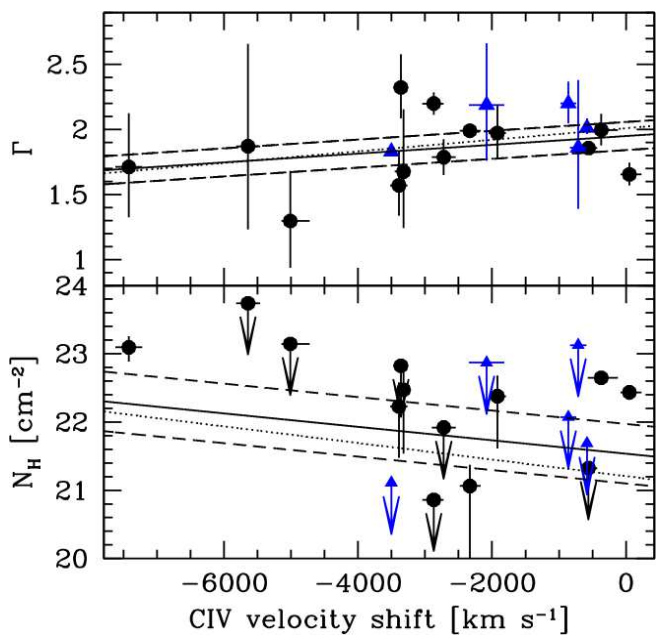}
   \end{center}
   \caption{$\Gamma$ and $\nhsym$ as a function of \vciv\ (top and bottom panels). The meaning of the data points and linear fits is the same as in Fig.~\ref{lxciv}.}
   \label{gnhshift}
\end{figure}

We notice that recent post-processing radiative transfer calculations suggest that the FW is not able to efficiently prevent over-ionisation and, therefore, it may not be so crucial for wind acceleration. Indeed, the FW is found to have a higher ionisation state (than previous estimates) and an extremely limited X-ray shielding power \citep{H2014}.  Evidence against the X-ray shielding scenario comes from X-ray observations of semi-relativistic BAL winds, reporting weak and moderate  X-ray absorption \citep[$\sim0.5-5\times10^{22}$~\nh; ][]{HC2013}. Similar evidence also seems to be present for hyper-luminous quasars. Indeed, our result would support a high degree of intrinsic X-ray quenching (Fig.~\ref{lxciv}, left panel) rather than a dependence on nuclear X-ray shielding (Fig.~\ref{gnhshift}, lower panel). 
 Further support for the hypothesis of coronal X-ray weakness comes from \nustar\ estimates that at least $\sim 1/3$ of luminous BAL quasars may exhibit significant X-ray weakness \citep{L2014}. However, the \citeauthor{L2014} sample consists of heavily obscured ($\nhsym\approx10^{24}$~\nh) BAL quasars, which, in principle, may lead to part of the X-ray emission being further suppressed by nuclear shielding.

Studies focusing on samples of weak-line quasars explain their
properties in the context of simple orientation-dependent nuclear
shielding of the X-ray emission without invoking coronal quenching
\citep{W2011,L2015,N2018}. In such a model, the shield may likely be
produced by the geometrically thick inner accretion disc regions,
and it would lead to a significant dependence on $\nhsym$ and a lack of
correlation with the intrinsic \lum. In our MIR and optically selected
WISSH quasars, we instead find a marked $\lumh$-\vciv\ dependence and a
lack of correlation with $\nhsym$. However, it is important to notice that only
simple cold absorbers have been adopted in modelling their X-ray spectrum thus far.  Probably the inner shielding disc regions would 
require a more complex and ionised absorber.
In this sense, the findings by \citet{HC2013} on semi-relativistic BAL winds in sources  with low and moderate observed cold $\nhsym$ may imply the existence of such an ionised absorber which would act as an ionisation shield.  
The current X-ray data quality for our sample is not sufficient enough to add further free
parameters to account for ionised absorption. 

We mention that orientation may also play a role in producing the large scatter in the $\lumh$-\vciv\ relation. Indeed it may reflect the projection of the wind velocity field depending on the line-of-sight inclination of the disc-corona structure. For instance, in the scenario envisaged by \citet{E2000}, which predicts an extreme funnel-shaped geometry of the wind for luminous quasars \citep[see Fig.~7 in ][]{E2000}, we qualitatively expect that at a given $\lumh$ the highest emission line blueshifts are reported by sources seen at relatively large inclinations (compatible with their Type~1 nature).\ Whereas, the lowest blueshifts are seen in sources viewed pole-on.

X-ray observations of well-monitored ultra-fast outflows \citep[UFOs; e.g. ][]{T2010,G2013} seem to support the 
  important role of radiation in driving AGN disc winds. For example, the hyper-luminous local quasar PDS~456, known to display recurrent and variable UFOs \citep[e.g.][]{Re2009,Na2015}, fits well in the  $\lumh$-\vciv\ relation as it is reported to have a CIV blueshift of $\sim5200~\kms$ \citep{OB2005} and $\lumh\approx(0.3-1)\times10^{45}$~\lumcgs. Interestingly, \citet{Ma2017} report a positively correlated variability on PDS~456 (over a period of 13 years) between the UFO velocity and the X-ray luminosity\footnote{We notice that variable UFOs have also been observed in other four high-z bright quasars \citep[APM~08279+5255, PG~1115+080, H~1413+117 and HS~1700+641;][]{Ch2003,Ch2007,SC2011,Lz2012}. A similar correlation between the quasar X-ray luminosity and the UFO velocity has been reported for APM~08279+5255 \citep{SC2011}.}. If PDS~456 is representative of the high-redshift hyper-luminous quasars, this level of X-ray variability could account for the scatter in the relation.   As for the opposite signs of correlation for CIV BEL winds and UFO velocities with $\lumh$, if proven to be a common characteristics in luminous quasars, they may as well be qualitatively explained in the context of the radiation-driven accretion disc wind scenario. Indeed, the X-ray luminosity, which acts on X-ray line transitions, is responsible for the radiative acceleration of the UFO in one case. In the other case, it acts as an ionisation state regulator of UV line-driven BEL winds. We notice though that recent photoionisation and radiative transfer calculations by \citet{D2019} suggest that the line driving mechanisms may not be relevant in plasma with high ionisation parameters typical of UFOs \citep[i.e. $\xi>1000$; e.g.][]{T2011,Na2015}.  In this case, MHD-driven winds may be a viable mechanism to explain the positive correlation exhibited by UFOs as recently reported by \citet{F2018}.

Dedicated deep X-ray observations, which will also be performed at lower luminosity regimes, will be crucial in testing and shedding light on the origin of the \vciv\ dependence by better constraining the spectral parameter for the X-ray weak sources. In this regard, {\it ATHENA} will further enable us to investigate the properties of the accretion disc-scale absorber (e.g. kinematics and ionisation state), discriminate between competing disc wind scenarios, and further investigate the role of UFOs \citep[e.g.][]{M2017,BC2019}.

Future studies on $\aox$, $k_{\rm bol,X}$, and $L_{6\mu m}/L_{2-10}$ will need to consider the
  reported marked dependences on  \vciv\ (Fig. ~\ref{lxciv} and ~\ref{lumsciv}) at these high luminosity regimes. This will be a necessary step in order to obtain better constrained relations and remove possible systematics due to the inclusion of highly blueshifted and hence X-ray weak sources.

\begin{acknowledgements}
We thank the referee for his/her comments and suggestions. 
We thank Silvia Martocchia for useful discussions.  LZ, EP, AB and MB acknowledge financial support under ASI/INAF contract 2017-14-H.0. 
CR acknowledges support from the CONICYT+PAI Convocatoria Nacional subvencion a instalacion en la academia convocatoria a\~{n}o 2017 PAI77170080. MG is supported by the ``Programa de Atracci\'on de Talento'' of the 
Comunidad de Madrid, grant number 2018-T1/TIC-11733.   GM is supported by the Spanish State Research Agency (AEI) Project No. ESP-2017-86582-C4-1-R. This research has been partially funded by the AEI Project No. MDM-2017-0737 Unidad de Excelencia “María de Maeztu” - Centro de Astrobiología (INTA-CSIC).

\end{acknowledgements}

\bibliographystyle{aa}
\bibliography{xwissh}

\begin{appendix} 
\section{Sample selection}\label{sampleselection}
\begin{figure}[t]
   \begin{center}
     \includegraphics[width=0.45\textwidth,angle=0]{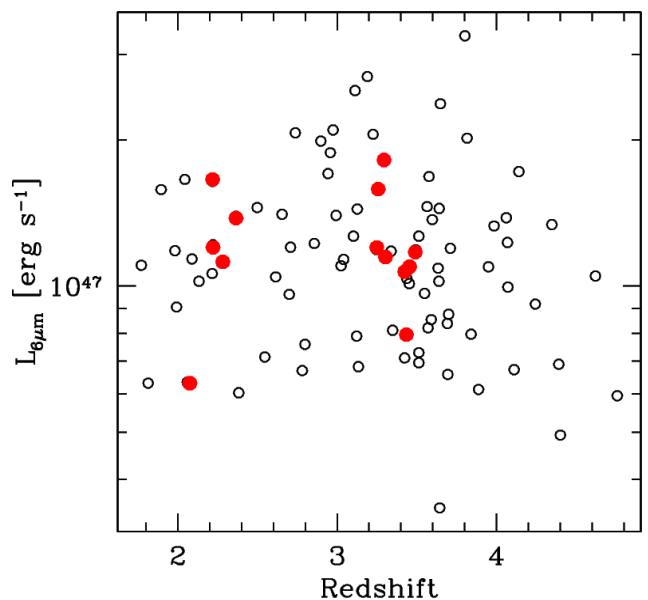}
   \end{center}
   \caption{$\lmirmon$ versus $z$ for our WISSH sub-sample (filled red circles) and the entire WISSH sample (empty black circles).}
   \label{lmirz}
\end{figure}

The sources considered in this work are reported in Table~\ref{wisshprop}. They were selected starting from the 18 WISSH quasars analysed by \citet{V2018} for which CIV emission line properties were estimated. We only include the twelve  radio-quiet sources with available archived X-ray \chandra\ and \xmm\ data (see Table~\ref{xraysample}). We added an additional WISSH source (J1441+0454) to this sample with available X-ray archived data for which constraints for the CIV emission line have already been obtained and will be reported in Vietri et al. in preparation. In Fig.~\ref{lmirz} we report the \lmir$-z$ for our sub-sample (red points) together with the entire WISSH sample. The sources in our sub-sample are mainly clustered at $z\approx2.1$ and $z\approx3.4$; therefore, they encompass the large $z$ range of the WISSH sources and exhibit \lmir\ values that are representative of the entire WISSH sample. 
X-ray data for eleven sources have already been analysed by \citet{M2017}.   In this work, additional and new proprietary and publicly available X-ray data are considered for seven sources (J0958+2827, J1201+0116, J1236+6554, J1326-0005, J0900+4215, J1106+6400, J2123-0050). Two of these sources (J0958+2827, J1326-0005) are not in the Martocchia et al. (2017) sample.
The final WISSH sample studied here consists of thirteen sources. \chandra\ and \xmm\ X-ray data are available for eleven and six sources, respectively. Three sources have both \chandra\ and \xmm\ coverage. The log of the X-ray observations for these sources is reported in Table~\ref{xraysample}.

In order to increase the number of hyper-luminous sources with available X-ray data and CIV spectral properties, we consider an additional sample starting from the sources analysed in C17 with $\lbol>10^{47}$~\lumcgs and available pointed X-ray data. The C17 sample is a heterogeneous compilation of 230 quasars with available infrared and optical spectroscopy. 
Our final sample consists of the five sources reported in Table~\ref{wisshprop}. \chandra\ and \xmm\ data are available for two and three sources, respectively (see Table~\ref{xraysample}).

Finally, the sources that are considered are all undetected  in the FIRST/NVSS radio surveys with the only exception of three sources (J0900+4215, J1441+0454, J1549+1245) with 1.4~GHz integrated flux densities of $\sim1.7~$mJy. From the integrated FIRST flux-densities at 1.4~GHz, we calculated the radio loudness parameter $R$ defined as the ratio between the rest-frame 6~cm and 2500~$\rm\AA$ flux densities\footnote{In order to derive the K-corrected rest-frame 6~cm flux densities from the 1.4~GHz observed ones, we assumed a power-law synchrotron spectrum $S(\nu)\propto\nu^{-\alpha}$ with $\alpha=0.5$ \citep[e.g.][]{Ji2007}.}. We obtained $R<10$ for all the sources and therefore we conclude that they are not radio-loud \citep{K1989}. 

\setlength{\tabcolsep}{7pt}
\begin{table*}
  \begin{center}
\caption{Properties of the WISSH quasars and the additional sources from the C17 sample}
  \label{wisshprop}
  \begin{tabular}{lcccccc}
    \hline
    \hline
    Name    &   z$^{a}$    &    \vciv$^{b}$        & $L_{\rm 2500~\AA}$$^{c}$ & $\nu  L_{\rm 6\mu m}$$^{c}$ & $\lbol$ & $\ledd$$^{d}$  \\
    (SDSS)  &        & ($\rm km~s^{-1}$)& $\rm log (erg~s^{-1}~Hz^{-1})$  & $\rm log (erg~s^{-1})$    & $\rm log (erg~s^{-1})$  &   \\
    \hline
    \multicolumn{7}{c}{WISSH sample}  \\
J0801+5210     & 3.257  &  $-2720 _{-180  }^{+180  }$  &  32.3  &  47.2  &  47.8$^{c}$  & $0.7 $     \\ 
J0900+4215$^e$ & 3.294  &  $-560  _{-110  }^{+130  }$  &  32.4  &  47.3  &  48.0$^{c}$  & $3.1 $     \\ 
J0958+2827     & 3.434  &  $-5640 _{-180  }^{+180  }$  &  32.1  &  46.9  &  47.6$^{c}$  & $0.8 $     \\ 
J1106+6400     & 2.221  &  $-2870 _{-170  }^{+150  }$  &  32.3  &  47.1  &  47.8$^{c}$  & $0.5 $     \\ 
J1111+1336     & 3.490  &  $-1920 _{-130  }^{+130  }$  &  32.2  &  47.1  &  47.7$^{c}$  & $0.5 $     \\ 
J1201+0116     & 3.248  &  $-3390 _{-130  }^{+130  }$  &  32.1  &  47.1  &  47.6$^{c}$  & $1.0 $     \\ 
J1236+6554     & 3.424  &  $-3360 _{-110  }^{+110  }$  &  32.2  &  47.0  &  47.7$^{c}$  & $0.8 $     \\ 
J1326-0005     & 3.303  &  $50    _{-130  }^{+180  }$  &  32.2  &  47.1  &  47.8$^{c}$  & $2.1 $     \\ 
J1421+4633     & 3.454  &  $-5010 _{-130  }^{+290  }$  &  32.2  &  47.0  &  47.7$^{c}$  & $0.6 $     \\ 
J1441+0454$^e$ & 2.075  &$-4210 _{-140  }^{+70   }$$^f$&  31.9  &  46.8  &  47.3$^{c}$  & $0.7$$^f$  \\ 
J1521+5202     & 2.218  &  $-7420 _{-200  }^{+200  }$  &  32.4  &  47.2  &  47.9$^{c}$  & $0.7 $     \\ 
J1549+1245$^e$ & 2.365  &  $-370  _{-200  }^{+250  }$  &  32.3  &  47.1  &  47.8$^{c}$  & $0.4 $     \\ 
J2123-0050     & 2.282  &  $-2330 _{-150  }^{+160  }$  &  32.2  &  47.0  &  47.7$^{c}$  & $1.1 $     \\ 
    \multicolumn{7}{c}{C17 sample}  \\
J0304-0008     & 3.296  &  $-583   \pm{16   }$       &  --    &  --    &  47.4$^g$     & $1.6 $    \\ 
J0929+2825     & 3.407  &  $-2080  \pm{264  }$       &  --    &  --    &  47.6$^g$     & $0.4 $    \\ 
J0942+0422     & 3.284  &  $-860   \pm{117  }$       &  --    &  --    &  47.8$^g$     & $1.5 $    \\ 
J1426+6025     & 3.197  &  $-3503  \pm{45   }$       &  --    &  --    &  48.0$^g$     & $0.5 $    \\ 
J1621-0042     & 3.729  &  $-713   \pm{124  }$       &  --    &  --    &  47.9$^g$     & $1.0 $    \\ 
    \hline
  \end{tabular}
  \end{center}
  \tablefoot{$^{a}$:  for the WISSH quasars, the redshifts were derived from the narrow component of the $H\beta$ emission line \citep[see][]{BP2017,V2018}; $^{b}$: peak of the CIV emission line in units of $\rm km~ s^{-1}$ relative to the $H\beta$ emission line as reported by \citet{V2018} and C17 for the WISSH and C17 samples, respectively. It is important to notice that \citet{V2018} and C17 adopt slightly different definitions of \vciv. \citet{V2018} estimated \vciv\ to be relative to the wavelength of the peak of the modelled CIV line, while C17 used the wavelength that bisects the cumulative flux distribution of the modelled line as a reference; $^{c}$: extinction-corrected derived from spectral energy distribution modelling (\citealt{D2017}; Duras et al. in prep.); $^{d}$: from \citet{V2018} and C17; $^{e}$: detected in the FIRST survey with 1.4~GHz integrated flux density of 1.7-1.8~mJy  ; $^{f}$: from Vietri et al. in prep.; $^{g}$: from $5100\rm\AA$ assuming a bolometric correction from \citet{R2012}.}
\end{table*}

\section{X-ray data reduction}\label{xrayreduction}
Data from sources listed in Table~\ref{xraysample} whose Observation ID (OBSID) is marked by an asterisk have been also analysed in \citet{M2017}. 
We performed the reduction of the new \chandra\ data with CIAO~4.9 (with version 4.7.3 of the \chandra\ Calibration Database). We reprocessed each data set with the script {\sc chandra\_repro} with the default script settings in order to generate level~2 data with the most updated calibration. We inspected the light-curves on each observation in order to check for possible high background periods and did not find any observation to be affected by such contaminations.  The spectral extraction and the creation of the relative response files  were performed using the CIAO script {\sc specextract}. We used source extraction circular regions with radii in the range of 1.5-3~arcsec in order to include all of the source counts. For the background, we adopted annular source-free regions centred on the quasar with inner and outer radii of 6 and 30~arcsec, respectively. For  the source j0958, we reduced two observations and then added spectral and response files with the FTOOLS script {\sc addascaspec}.

The reduction of the \xmm\ data (both pn and MOS cameras\footnote{Compared to \citet{M2017} who analysed pn-only spectra, we further improved the statistics of the \xmm\ spectra by adding the MOS data.}) was performed with SAS v16.0.0. All of the observations were performed in full-window mode with the thin filter applied. The light curves for pn and MOS exposures were screened at energies $> 10$~keV ($10-12$~keV for the pn) for high background flaring periods. We adopted a count-rate threshold filtering criterion with typical values of 0.3-0.5 and 0.1-0.2 counts~$s^{-1}$ for pn and MOS, respectively.  The resulting net-exposure times are reported in Table~\ref{xraysample}. We selected X-ray events corresponding to patterns 0-4 and 0-12 for pn and MOS, respectively. The source extraction was performed using the same circular apertures for both pn and MOS detectors. In order to include all of the source counts and simultaneously minimise the background contribution for each source, we adopted different apertures ranging from 12 to 30 arcsec. The background spectrum was extracted in the chip including the source from circular (1-2~arcmin radius) and annular (inner and outer radii up to 0.7 and 4 arcmin) source-free regions for pn and MOS, respectively.
The same data reduction steps were followed in case of the sources belonging to the C17 sample (see Table~\ref{xraysample} for details on their X-ray observations).

For \chandra\ spectra, we obtained background-subtracted counts ranging from 24 to $\sim200$ in the 0.3-8~keV band, reaching $\sim800$~counts in one case (J2123$-$0050). As for  \xmm,\  we obtained spectra with background-subtracted counts ranging from 70 (126) to 1700 (1600) for pn (MOS1+MOS2) detectors. 
We consistently grouped all \chandra\ and \xmm\ spectra at a minimum of one background-subtracted count per bin. The X-ray modelling was performed with XSPEC v.12.9.0i by adopting the Cash statistic (C-stat) with the implemented direct background subtraction \citep{C1979,WLK1979}.

\setlength{\tabcolsep}{3pt}
\begin{table*}[!t]
  \begin{center}
  \caption{Log of the X-ray observations of the sources from the WISSH and C17 samples}    
  \label{xraysample}
  \begin{tabular}{lcccccccc}
      \hline
      \hline
Name          &  \multicolumn{4}{|c|}{\chandra/ACIS-S}          &    \multicolumn{4}{c}{\xmm}                                    \\
(SDSS)        &\multicolumn{1}{|c}{OBSID$^b$}       & Date       & Exposure$^{c}$& \multicolumn{1}{c|}{Counts$^{d}$}   &OBSID$^b$      &   Date       &          Exposures$^{c}$    & Counts$^{d}$           \\
\hline
    \multicolumn{9}{c}{WISSH sample}  \\
J0801+5210    & 17081$^*$   & 2014-12-11 &   43.5  &    179   &   -        &      -       &      -          &         -           \\
J0900+4215    &  6810$^*$   & 2006-02-09 &    3.9  &    118   & 0803950601 & 2017-11-17   &    12.8/ 18.8/ 18.5  &   1710/  607/  663  \\
J0958+2827    &20444/21863$^\dagger$ & 2018-09-30 & 40.8 &  24 &   -        &      -       &      -          &         -           \\
J1106+6400    &  6811$^*$   & 2006-07-16 &    3.6  &    142   & 0553561401 & 2008-11-29   &     1.2/  5.0/  5.1  &     67/   85/   79       \\
J1111+1336    &  17082$^*$  & 2015-01-26 &   43.1  &    184   &   -        &      -       &      -          &         -           \\
J1201+0116    &   -         &   -        &     -   &     -    & 0803952201 & 2017-06-06   &    34.8/ 40.9/ 41.1  &    154/   70/   56  \\
J1236+6554    &  20443$^\dagger$ & 2018-07-29 & 44.9 &   138   &   -        &      -       &      -          &         -           \\
J1326$-$0005  &   -         &   -        &     -   &     -    & 0804480101$^\dagger$ & 2017-12-30  &  34.8/ 43.7/ 44.8 &    438/  169/  196  \\
J1421+4633    &  12859$^*$  & 2011-06-20 &   23.6  &     51   &   -        &      -       &      -          &         -           \\
J1441+0454    &  12860$^*$  & 2011-12-14 &   21.5  &     79   &   -        &      -       &      -          &         -           \\
J1521+5202    &  15334      & 2013-10-22 &   37.4  &     83   &   -        &      -       &      -          &         -           \\
J1549+1245    &   -         &   -        &     -   &     -    & 0763160201$^\dagger$ & 2016-02-04  &  28.1/ 57.1/ 43.6 &    504/  254/  278  \\
J2123$-$0050  &   17080$^\dagger$ & 2015-12-22 & 39.6  &  774  & 0745010401$^\dagger$ & 2014-11-14   &    19.7/ 28.2/ 29.4  &    552/  158/  162  \\
\multicolumn{9}{c}{C17 sample}  \\
J0304-0008    & -     &    -       &    -    &  -        & 0803952901  & 2017-08-26   &    30.4/ 36.2/ 36.2  &    840/  259/  244   \\
J0929+2825    & 10740 & 2009-01-07 &    5.0  &     35    &     -       &      -       &           -          &           -          \\
J0942+0422    & -     &    -       &    -    &  -        & 0803951801  & 2017-04-30   &     9.6/ 22.1/ 18.0  &    218/   87/  115   \\
J1426+6025    & -     &    -       &    -    &  -        & 0803950301  & 2017-05-12   &    18.4/ 23.4/ 23.2  &   1223/  310/  368   \\
J1621-0042    & 2184  & 2001-09-05 &    1.6  &     27    &     -       &      -       &           -          &           -          \\
\hline
  \end{tabular}
  \end{center}
  \tablefoot{$^{a}$: observation ID;  $^b$: cleaned exposure time in units of ksec; $^{c}$: background-subtracted counts on the 0.3-8~keV and 0.2-10~keV bands for \chandra\ and \xmm, respectively; $^*$: data reduced in Martocchia+2017; $^\dagger$: new proprietary data}
\end{table*}

\section{X-ray spectral analysis}\label{xrayanalysis}
The X-ray modelling, which was consistently carried out with the same model for all of the sources, was performed in the 0.2-10~keV and 0.3-8~keV for the \xmm\ and \chandra\ detectors, respectively. The adopted model is a simple power-law model absorbed by the Galactic interstellar medium and by the obscuring medium (i.e. nuclear absorber+interstellar matter) at the redshift of the source. The obscuration is parametrised as uniform cold absorbers adopting the {\sc wabs} and {\sc zwabs} multiplicative models for Galactic and intrinsic  absorbers, respectively. In the cases of sources with spectra from  more than one detector (i.e. either for \xmm\ or \xmm+\chandra), a joint fit is performed with the addition in the modelling of a constant term to account for cross-calibration and possible source flux variation (in the case of observations taken at different epochs). During the modelling, we tied the pn and MOS constants whenever the MOS one was exceeded by more than 7\% the pn constant\footnote{The MOS detectors have been reported to have at most 7\% higher fluxes than the pn as detailed in the XMM-SOC-CAL-TN-0052 Issue 6.0 available in http://xmm2.esac.esa.int/docs/documents/CAL-TN-0052.ps.gz.}. For all of the sources  except J0900+4215, the calibration constants have been linked among the \xmm\ detectors. For J0900+4215 (the source with the best quality spectra), the MOS spectra were found to have constants $\sim3\%$ higher than the PN. For three sources, a joint fit analysis with \xmm\ and \chandra\ was performed (i.e. J0900+4215, J1106+6400, J2123$-$0050). We find J0900+4215 and J1106+6400 to exhibit \chandra\ calibration constants, which significantly differ from the \xmm\ ones by a factor $\sim0.6\pm0.1$ and $1.8\pm0.3$ (errors are 90\% level), respectively. This indicates that the sources have varied their flux between the two observations. 
For J2123$-$0050, the \chandra\ constant is a factor $>2$ higher than the \xmm\ ones. This is likely due to flux loss in the \xmm\ spectra due to the presence  of a source at $\sim20$~arcsec from the quasar, which forced us to shrink the quasar spectral extraction region to a circle with a radius of 12~arcsec in order to minimise the contamination. 
Apart from the calibration constants, the modelling was performed by linking all of the other parameters and leaving three parameters free to vary, that is, the source column density ($\nhsym$), photon index ($\Gamma$), and power-law normalisation. Error estimation for $\nhsym$\ and $\Gamma$ was performed by leaving all of the previously discussed parameters free to vary during the calculation. The uncertainty on the unabsorbed X-ray luminosities (i.e. 2~keV and 2-10~keV) were estimated by freezing $\Gamma$ to its best-fit value. 
For each source, Table~\ref{wisshxray} reports the fit statistic (C-stat) and degrees of freedom (dof) of the modelling along with $\Gamma$, $\nhsym$, the observed fluxes at 0.5-2~keV and 2-10~keV ($f_{0.5-2}$ and $f_{2-10,}$ respectively), $\lumh$, and the corrected (for extinction and absorption) $\aox$ and $\Delta\aox$ (based on \citealt{J2007}).  Determinations of $\aox$ from the literature are usually derived based on 2~keV luminosity estimated under the assumption of a power-law X-ray spectral model without accounting for intrinsic absorption. Hence, for consistency, we also derived $\aox^{pow}$ and $\Delta\aox^{pow}$ by estimating the 2~keV luminosity from simple unabsorbed power-law spectral modelling (we only accounted for Galactic absorption). 
As for the quantities regarding fluxes and luminosities (i.e. $f_{0.5-2}$, $f_{2-10}$, $\lumh$,  $\aox,$ and $\Delta\aox$), we adopted and report the pn-derived values for all of the sources observed by \xmm-only in Table~\ref{wisshxray}. For the sources for which we performed joint \xmm\ and \chandra\ modellings, we adopted the \xmm-derived quantities for J1106+6400 (the \chandra\ observations are of lower quality) and the \chandra\ ones for J0900+4215 and J2123$-$0050. Indeed, for J0900+4215, the \chandra\ observation was performed closest in time to the SDSS spectrum in which the relative \vciv\ was measured. For J2123$-$0050, the \chandra\ spectrum does not suffer from contamination from the nearby X-ray source. 

We tested for possible X-ray model-dependent systematics on the derived $\lumh$. For each WISSH source with best-fit values or upper limits on $\nhsym$ larger than $10^{22}$~\nh, we parametrised the intrinsic absorption  with a partially ionised absorber by adopting the model {\sc zxipcf} with the covering fraction $f_c=1$. We left both $\nhsym$ and the ionisation parameter ($\xi$) free to vary. We obtain an estimated $\lumh,$ which is on average 0.15~dex larger than the luminosities derived by using a cold absorption model. In this case, we still have a significant $\lumh$-\vciv\ relation ($\rho=0.74$ and $p=0.006$).  

We mention that restricting the energy range of the WISSH spectral modellings by limiting the low energy bound to 0.5~keV for all the instruments does not make a significant difference in our result. Indeed, $\lumh$ does not vary by more than $\pm0.05~dex$. This level of variation does not change the strength of the reported correlation with \vciv\ (i.e. $\rho=0.75$ and $p=0.004$). 

\setlength{\tabcolsep}{3pt}
\begin{table*}[!t]
  \begin{center}
  \caption{Properties of the WISSH and C17 samples derived from the X-ray spectral analysis}    
  \label{wisshxray}
  \begin{tabular}{lcccccccrr}
      \hline
      \hline
Name   & C-stat &  dof  &          $\Gamma$    & $\log N_{\rm H}$    &  $\log f_{0.5-2}$$^a$          &  $\log f_{2-10}$$^b$        &  $\log \lumh$$^c$       &   $\aox(\aox^{pow})$$^d$ & $\Delta\aox(\Delta\aox^{pow})$$^e$   \\        
(SDSS) &        &       &                      &  $\log(\nhi)$                  &                               &                            &  $\log$(\lumcgs)           &             &           \\ 
\hline
\multicolumn{10}{c}{WISSH sample}  \\
J0801+5210    &   90.5 &    87 & $1.79_{-0.14}^{+0.14}$ & $<21.9                 $ &            1.5 &  2.7                       & $45.30_{-0.03}^{+0.03}  $   &  $-2.01_{-0.01}^{+0.01}$   $(-2.01)$          & $-0.20$  $(-0.20)$       \\ 
J0900+4215    & 1014.8 &  1089 & $1.86_{-0.03}^{+0.04}$ & $<21.3                 $ &            8.5 & 13.1                       & $46.04_{-0.01}^{+0.01}  $   &  $-1.76_{-0.00}^{+0.01}$   $(-1.76)$          & $+0.07$  $(0.07)$        \\ 
J0958+2827    &    8.5 &     9 & $1.87_{-0.64}^{+0.79}$ & $<23.7                 $ &            0.2 &  0.6                       & $44.73_{-0.13}^{+0.13}  $   &  $-2.14_{-0.05}^{+0.05}$   $(-2.27)$          & $-0.36$  $(-0.48)$       \\ 
J1106+6400    &  208.7 &   203 & $2.20_{-0.08}^{+0.09}$ & $<20.9                 $ &            6.5 &  5.8                       & $45.46_{-0.03}^{+0.03}  $   &  $-1.89_{-0.01}^{+0.01}$   $(-1.89)$          & $-0.08$  $(-0.08)$       \\ 
J1111+1336    &   66.5 &    95 & $1.97_{-0.20}^{+0.21}$ & $22.4_{   -0.8}^{+    0.3}$ &            1.6 &  2.4                       & $45.45_{-0.04}^{+0.04}  $   &  $-1.91_{-0.08}^{+0.08}$   $(-1.96)$          & $-0.10$  $(-0.16)$    \\
J1201+0116    &  128.4 &   178 & $1.57_{-0.23}^{+0.27}$ & $22.2_{   -0.8}^{+    0.4}$ &            0.6 &  1.5                      & $44.90_{-0.05}^{+0.05}  $   &  $-2.14_{-0.02}^{+0.02}$   $(-2.14)$          & $-0.35$  $(-0.35)$    \\    
J1236+6554    &   54.8 &    56 & $2.32_{-0.24}^{+0.25}$ & $<22.8                 $ &            1.7 &  1.5                       & $45.45_{-0.05}^{+0.05}  $   &  $-1.85_{-0.02}^{+0.02}$   $(-1.89)$          & $-0.05$  $(-0.09)$       \\ 
J1326$-$0005  &  425.4 &   403 & $1.66_{-0.09}^{+0.09}$ & $22.4_{   -0.1}^{+    0.1}$ &            1.7 &  4.3                          & $45.43_{-0.02}^{+0.02}  $   &  $-1.96_{-0.01}^{+0.01}$   $(-1.96)$          & $-0.16$  $(-0.16)$    \\  
J1421+4633    &   19.7 &    28 & $1.30_{-0.36}^{+0.39}$$^g$ & $<23.1                 $ &            0.6 &  2.5                       & $45.00_{-0.08}^{+0.07}  $   &  $-2.16_{-0.03}^{+0.03}$   $(-2.24)$          & $-0.36$  $(-0.44)$       \\ 
J1441+0454    &   49.1 &    45 & $1.68_{-0.43}^{+0.48}$ & $22.5_{   -0.9}^{+    0.4}$ &            1.0 &  3.1                          & $44.87_{-0.07}^{+0.07}  $   &  $-2.04_{-0.03}^{+0.03}$   $(-2.15)$          & $-0.28$  $(-0.39)$     \\
J1521+5202    &   63.8 &    40 & $1.71_{-0.39}^{+0.41}$ & $23.1_{   -0.2}^{+    0.2}$ &            0.5 &  2.8                          & $44.93_{-0.07}^{+0.07}  $   &  $-2.21_{-0.03}^{+0.03}$   $(-2.50)$          & $-0.38$  $(-0.67)$     \\   
J1549+1245    &  480.6 &   546 & $2.00_{-0.12}^{+0.12}$ & $22.6_{   -0.1}^{+    0.1}$ &            2.2 &  4.8                          & $45.35_{-0.02}^{+0.02}  $   &  $-1.96_{-0.01}^{+0.01}$   $(-1.96)$          & $-0.15$  $(-0.15)$     \\   
J2123$-$0050  &  613.3 &   565 & $1.99_{-0.06}^{+0.06}$ & $21.1_{   -1.9}^{+    0.3}$ &            8.9 & 11.8                          & $45.69_{-0.02}^{+0.02}  $   &  $-1.80_{-0.01}^{+0.01}$   $(-1.80)$          & $+0.00$  $(-0.00)$     \\ 
\multicolumn{10}{c}{C17 sample}  \\
J0304$-$0008  &  532.7 &   586 & $2.01_{-0.05}^{+0.07}$ & $<21.7$                  &                3.5 &  4.8                           & $45.70_{-0.02}^{+0.02}$ &  \multicolumn{1}{c}{--} &  \multicolumn{1}{c}{--} \\
J0929$+$2825  &  123.2 &   523 & $2.19_{-0.43}^{+0.48}$ & $<22.9$                  &                2.3 &  2.5                           & $45.57_{-0.09}^{+0.09}$ &  \multicolumn{1}{c}{--} &  \multicolumn{1}{c}{--} \\
J0942$+$0422  &  219.1 &   224 & $2.20_{-0.15}^{+0.17}$ & $<22.1$                  &                2.5 &  2.5                           & $45.53_{-0.04}^{+0.04}$ &  \multicolumn{1}{c}{--} &  \multicolumn{1}{c}{--} \\
J1426$+$6025  &  645.1 &   740 & $1.83_{-0.04}^{+0.04}$ & $<21.1$                  &                6.5 & 10.4                           & $45.89_{-0.01}^{+0.02}$ &  \multicolumn{1}{c}{--} &  \multicolumn{1}{c}{--} \\
J1621$-$0042  &    9.5 &    12 & $1.86_{-0.47}^{+0.52}$ & $<23.1$                  &                4.1 &  8.6                           & $45.99_{-0.11}^{+0.10}$ &  \multicolumn{1}{c}{--} &  \multicolumn{1}{c}{--} \\
  \hline
  \end{tabular}
  \end{center}
  \tablefoot{$^{a}$: 0.5-2~keV observed flux in units of $10^{-14}$~\fluxcgs;  $^b$: 2-10~keV observed flux in units of $10^{-14}$~\fluxcgs; $^{c}$: 2-10~keV unabsorbed luminosity; $^d$: $\aox$ computed using extinction-corrected 2500~$\rm\AA$ and unabsorbed 2~keV luminosities. The $\aox^{pow}$ with 2~keV luminosity from unabsorbed power-law modelling is reported in parenthesis.; $^e$: difference between the measured $\aox$ and predicted one based on the $\aox-L_{\rm 2500~\rm\AA}$ relation from \citet{J2007}. In this luminosity regime, the recently derived $\aox-L_{\rm 2500~\rm\AA}$ relation by \citet{LR2016}, which was derived for the X-ray detected quasars with $E(B-V)<0.1$ and $1.6\leq\Gamma\leq 2.8$, predict $\aox$ values $\sim0.03$ smaller (i.e. more negative) than the \citet{J2007} ones. In parenthesis, $\Delta\aox^{pow}$ that adopts $\aox^{pow}$ with 2~keV luminosity from unabsorbed power-law modelling is reported; $^g$: For this source the flat $\Gamma$ value is consistent with the canonical $\Gamma=1.8-2$ derived for local PG quasars \citep{P2015}. We verified that the inferred low luminosity is not the result of the flat best-fit photon index. Hence, we fitted the data with $\Gamma=1.9$ and obtained a luminosity value, which is only 0.2~dex higher than estimated, therefore confirming its low X-ray luminosity.}  
\end{table*}

\end{appendix}
\end{document}